\documentclass[a4paper, oneside, reqno]{amsart}

\usepackage[dvips]{graphicx}
\usepackage{amsmath}
\usepackage{amssymb}
\usepackage{amsfonts}
\usepackage[mathscr]{eucal}
\usepackage{hyperref}  

\newcommand{\eref}[1]{(\ref{#1})}
\newcommand{\bs}[1]{\boldsymbol{#1}}

\begin{document}

\title[The Wigner-Weyl formalism versus complex geometrical optics]{The relationship between the Wigner-Weyl kinetic formalism and the complex geometrical optics method}
\author{Omar Maj}
\address{Physics Department ``A.~Volta'', University of Pavia, I-27100 Pavia, Italy}
\email{maj@fisicavolta.unipv.it}

\begin{abstract}
The relationship between two different asymptotic techniques, namely, the Wigner-Weyl kinetic formalism and the complex geometrical optics method, is addressed within the framework of semiclassical theory of wave propagation. More specifically, in correspondence to appropriate boundary conditions, the solution of the wave kinetic equation, relevant to the Wigner-Weyl formalism, is obtained in terms of the corresponding solution of the complex geometrical optics equations. In particular, this implies that the two considered techniques yield the same wavefield intensity. Such a result is also discussed on the basis of the analytical solution of the wave kinetic equation specific to Gaussian beams of electromagnetic waves propagating in a ``lens-like'' medium for which the complex geometrical optics solution is already available.      
\end{abstract}

\maketitle

\section{Introduction}
In the framework of the semiclassical theory of waves [1-5], i.e., short wavelength asymptotics, the uniform (global) description of the wavefield is complicated by the formation of caustic singularities [1, 2, 6-8]. Although a complete and deep understanding of the wavefield structure near caustic regions is obtained on the basis of catastrophe theory \cite{Dui, KO_Caustics, Nye} and the unfolding of the corresponding singularities can be treated by means of symplectic techniques \cite{GS1, Dui, GS2}, the application of such methods to realistic cases, e.g., to waves in magnetically confined plasmas \cite{STIX}, appears rather difficult. Therefore, with specific regard to physical applications, several asymptotic methods have been developed which yield numerically tractable equations, though being limited concerning the global properties of the asymptotic solutions.   

Such asymptotic techniques can be classified into two different families, depending on whether the relevant wave equation is described in the phase space, \emph{microlocal techniques}, or directly in the configuration space where the wavefield is defined, \emph{quasi-optical methods}.

This work aims to give a detailed comparative analysis of two such techniques, namely, the Wigner-Weyl kinetic formalism [11-13] and the complex geometrical optics (CGO) method [14-17] which can be considered as benchmarks for microlocal and quasi-optical methods, respectively. 

Specifically, in Sec.2, the Wigner-Weyl formalism and the complex geometrical optics method are reviewed and compared. In particular, it is pointed out that, within the Wigner-Weyl formalism, physically meaningful solutions should have a specific form, referred to as momentum distribution, which is characterized in Sec.3.   

On the basis of the mathematical properties of momentum distributions, our main result is obtained in Sec.4. In particular, it is shown that, in correspondence to appropriate boundary conditions, there exists a specific asymptotic solution of the wave kinetic equation relevant to the Wigner-Weyl formalism that can be written in terms of the corresponding solution of the complex geometrical optics equations. This allows us to relate the two considered methods as well as to determine the specific class of boundary conditions for which they are equivalent. In Sec.5, this general result is illustrated by means of an analytically tractable example, i.e., the propagation of a Gaussian beam of electromagnetic waves in an isotropic ``lens-like'' medium. In conclusion, a summary of the main results is given in Sec.6.

\section{The Wigner-Weyl Formalism versus the CGO method}

In order to set up the framework, let us discuss the relevant boundary value problem for a generic \emph{scalar} pseudodifferential wave equation, together with the required mathematical definitions. 

Specifically, we will consider the case of a scalar (real or complex) wavefield $\psi(x)$ propagating in the $N$-dimensional linear space $\mathbb{R}^{N}$ with $x=(x^1,\ldots,x^N)$ a generic set of Cartesian coordinates and denote by $k=(k_1,\ldots,k_N)$ the corresponding coordinates in the dual space $(\mathbb{R}^N)'\cong  \mathbb{R}^N$. Time-dependent wavefields are included as one of the coordinates can play the role of time, e.g., $x^N =ct$, $c$ being a reference speed, and the corresponding dual coordinate is related to frequency, e.g., $k_N =-\omega/c$. To some extent, the results for a scalar wave equation are valid also for a multi-component wave equation as the latter can be reduced to a set of \emph{independent} scalar equations far from mode conversion regions \cite{LF}.

Thereafter, the Wigner-Weyl formalism will be formulated entirely in the space $\mathbb{R}^N \times (\mathbb{R}^N)' \cong \mathbb{R}^{2N}$, with coordinates $(x,k)$, which is viewed as the trivial cotangent bundle \cite{Ht} over the configuration space $\mathbb{R}^N$ where the wavefield is defined. No explicit reference to the propagation direction is made, differently from the classical derivations \cite{McD}. As for the CGO method, it has been originally developed for solving the second-order partial differential equation relevant to the propagation of electromagnetic wave beams in stationary spatially nondispersive media. Hence, we will need to discuss its application to generic pseudodifferential equations. In particular, it is shown that the CGO method yields an approximation of the wavefield directly in the configuration space, provided that the wave equation satisfies an appropriate condition.

First, let us define the class of wave equations undergone to solution. A pseudodifferential wave equation is an equation of the form
\begin{equation}
\label{1n}
\big(\hat{D}\psi\big)(x)=\frac{1}{(2\pi)^N} \int e^{i k\cdot (x-x')} d(x,x',k) \psi(x')d^Nx'd^Nk =0,
\end{equation}
which admits propagating wave solutions. The operator $\hat{D}$ is a pseudodifferential operator [3,18-20] acting on the wavefield as a Fourier integral operator \cite{FIO} characterized by the bilinear phase function $k \cdot (x-x')$. Here, $d(x,x',k)$ belongs to a particular class of smooth functions, referred to as \emph{symbols}, which, roughly speaking, behave like a polynomial in $k$ for $|k|$ large enough. Specifically, a smooth function $a(z,k)$ with $z\in \mathbb{R}^M$ and $k\in (\mathbb{R}^N)'$ is a symbol of order $m\in \mathbb{R}$ if for every multi-indices $\alpha$, $\beta$ there is a constant $C_{\alpha,\beta}>0$ such that
\begin{equation}
\label{2n}
|\partial^{\alpha}_z \partial^{\beta}_k a(z,k)| \leq \frac{C_{\alpha,\beta}}{L^{|\alpha|-|\beta|}} \big( 1 + |kL| \big)^{m-|\beta|},\qquad \text{uniformly in $(z,k) \in \mathbb{R}^{M} \times (\mathbb{R}^N)'$},
\end{equation}
and one writes $a \in S^m(\mathbb{R}^{M} \times (\mathbb{R}^N)')$. In virtue of the symbol estimate \eref{2n} the integral in \eref{1n} makes sense for $\psi \in \mathcal{S}'(\mathbb{R}^{N})$, the space of tempered distribution \cite{T, Ht}. Moreover, the scale length $L$ characterizes the variations of symbols with respect to the spatial coordinate $z$ and it can be eliminated by the rescaling $z \to z/L$ and $k \to kL$. It is worth noting that any linear differential operator with smooth and bounded coefficients is a pseudodifferential operator \cite{H}.
 
Boundary conditions of Cauchy type are given on an $(N-1)$-dimensional hypersurface $\Sigma$: for simplicity, one can assume $\Sigma$ to be the hyperplane $\{x : x^N=0\}$ where the wavefield $\psi_0(\bs{x})$, $\bs{x}=(x^{1},\ldots,x^{N-1})$, is assigned together with as many derivatives $\psi_n(\bs{x}) = \partial^n \psi /\partial (x^N)^n$ as appropriate. We are interested in semiclassical solutions for which only the covectors $k$ with
\begin{equation*}
\kappa = |kL| \gg 1,
\end{equation*}
are significant, in the integral in \eref{1n}. As a consequence, the wavefield should be a highly oscillating function on the (large) scale length $L$ and it should correspond to a specific set of highly oscillating boundary conditions of the form
\begin{equation}
\label{1nbound}
\psi_0(\bs{x}) = A_0(\bs{x}) e^{iS_0(\bs{x})},\qquad \partial_{\bs{x}} S_0(\bs{x})L \sim \kappa \gg 1,
\end{equation}
the amplitude $A$ being slowly varying, that is, $|\partial_x A | \sim |A|/L$.

\subsection{The Wigner-Weyl formalism}
In the Wigner-Weyl formalism, the Weyl-symbol map $\sigma^W$ is applied in order to represent the wave equation \eref{1n} in $\mathbb{R}^{N} \times (\mathbb{R}^N)'$ which is naturally endowed with a phase-space structure. The Weyl-symbol map transforms an operator $\hat{A} : \mathcal{S}(\mathbb{R}^N) \to \mathcal{S}'(\mathbb{R}^{N})$,  $\mathcal{S}(\mathbb{R}^{N})$ being the space of Schwartz's functions \cite{Ht}, into a tempered distribution by acting on the Schwartz kernel $\mathscr{A}(x_1,x_2) \in \mathcal{S}'(\mathbb{R}^{2N})$ of the operator $\hat{A}$ according to \cite{McD}
\begin{equation}
\label{3n}
\mathscr{A}(x_1,x_2) \mapsto \sigma^W (\hat{A})(x,k) = \int \mathscr{A}\big(x+\tfrac{1}{2}s,x-\tfrac{1}{2}s\big) e^{-ik \cdot s} d^Ns.
\end{equation}
It is worth noting that for the pseudodifferential operator in \eref{1n} the Schwartz kernel is
\begin{equation*}
\mathscr{D}(x_1,x_2) = \frac{1}{(2\pi)^N} \int e^{ik \cdot(x_1-x_2)} d(x_1,x_2,k) d^Nk,
\end{equation*}
thus, the image of $\hat{D}$ under the Weyl-symbol map amounts to the formal series of decreasing order symbols
\begin{equation*}      
\sigma^W(\hat{D})(x,k) = \sum_\alpha \frac{i^{|\alpha|}}{\alpha!} \partial_s^\alpha \partial_k^\alpha d\big(x+\tfrac{1}{2}s,x-\tfrac{1}{2}s,k\big)\big|_{s=0},
\end{equation*}
where Taylor expansion has been used and $\partial_s^\alpha \partial_k^\alpha d \in S^{m-|\alpha|}$. Series of that kind admit always an asymptotic resummation \cite{Mar, H} to a symbol of order $m$, 
\begin{equation}
\label{weyl}
D(x,k) \sim \sigma^W(\hat{D})(x,k) \in S^m,
\end{equation}
which is referred to as \emph{Weyl symbol} of $\hat{D}$. In \eref{weyl} the $\sim$ denotes the asymptotic equivalence of symbols \cite{Mar, H}. On the other hand, one can consider the correlation operator \cite{McD} $\hat{S}$ whose Schwartz kernel is given by the tensor product $\psi(x_1)\psi^*(x_2)$, then the Weyl-symbol map yields the \emph{Wigner function} 
\begin{equation}
\label{wigner}
W(x,k) = \sigma^W(\hat{S})(x,k) = \int \psi\big(x+\tfrac{1}{2}s\big)\psi^*\big(x-\tfrac{1}{2}s)e^{-ik\cdot s} d^Ns.
\end{equation}

The wave equation \eref{1n} can be written in the equivalent form $\hat{D} \hat{S}=0$ and, on applying the Weyl-symbol map, one gets \cite{Mar, McD}
\begin{equation}
\sigma^W(\hat{D} \hat{S})(x,k) \sim \sum_{\alpha,\beta} \frac{(-1)^{|\alpha|}}{(2i)^{|\alpha + \beta|} \alpha!\beta!} \big(\partial_x^\alpha\partial_k^\beta D(x,k)\big)\big(\partial_x^\beta\partial_k^\alpha W(x,k)\big) =0.
\end{equation}
In the semiclassical limit one has $\partial_x^\alpha \partial_k^\beta D =O(\kappa^{m-|\beta|})$ so that, on assuming the same ordering for the Wigner function \cite{McD}, the foregoing equation separates into
\begin{subequations}
\label{1.1}
\begin{align}
\label{1.1a}
&D'(x,k) W(x,k)=0,\\
\label{1.1b}
&\big\{W(x,k),D'(x,k)\big\} = 2D''(x,k) W(x,k),
\end{align}
\end{subequations}
where $D' \in S^m$ and $D'' \in S^{m-1}$ are the real and imaginary parts of the Weyl symbol; in particular, one has $D''/D' = O(\kappa^{-1})$ which is the condition for weak absorption and/or instabilities \cite{McD}. 
Equation \eref{1.1a} is a constraint to Eq.\eref{1.1b} which, on the other hand, has the form of a kinetic equation in the $x$-$k$ phase space, $\{\cdot,\cdot\}$ being the corresponding Poisson brackets. We will refer to the whole system \eref{1.1} as the \emph{wave kinetic equation}.

In general, a solution of the wave kinetic equation is a tempered distribution, however, one usually restricts the class of solutions to semiclassical measures. This allows us to make sense of the integrals of the form \cite{Mar, McD}
\begin{equation*}
\langle \hat{A} \rangle_{\psi} = \frac{1}{(2\pi)^N} \int A(x,k)W(x,k)d^Nxd^Nk,
\end{equation*}
which expresses the expectation value of a physical quantity represented by the pseudodifferential operator $\hat{A}$ with Weyl symbol $A(x,k)$. In the following we will assume further regularity with respect to $x$ so that the expectation values of physical quantities can be defined \emph{locally}, that is,
\begin{subequations}
\label{physics}
\begin{equation}
\label{star}
A(x) =  \int\frac{d^Nk}{(2\pi)^N}\ A(x,k)W(x,k)
\end{equation}
makes sense as a smooth function in $x\in\mathbb{R}^N$. In particular, the wavefield intensity amounts to
\begin{equation}
\label{starstar}
|\psi(x)|^2 =  \int\frac{d^Nk}{(2\pi)^N}\ W(x,k).
\end{equation}
\end{subequations}
Such a restriction of the class of solutions is justified as, in general wave propagation problems, one needs a description of the space and time profiles of physical quantities.

Let us now discuss the appropriate boundary conditions for \eref{1.1}. On the hyperplane $\Sigma = \{x: x^N=0\}$ the wavefield $\psi_0(\bs{x})$ has been assigned and one can compute the corresponding Wigner function $W_0(\bs{x},\bs{k}) = \sigma^W(\hat{S}_0)(\bs{x},\bs{k})$, $\hat{S}_0$ being the correlation operator associated to $\psi_0$ and $\bs{k}=(k_1,\ldots,k_{N-1})$ the coordinates dual to $\bs{x}$; then a solution $W(x,k)$ should match $W_0(\bs{x},\bs{k})$ in some appropriate sense. Specifically, one should impose that the local value $A(x)$ of any physical quantity evaluated on $\Sigma$ is the same whether it is evaluated by $W(x,k)$ or by $W_0(\bs{x},\bs{k})$. Within this formulation, the Weyl symbol $A(x,k)$ should be restricted to $\mathbb{R}^{2N-2}$ where $W_0$ is defined. One can note that the suitable embeddings $\Gamma$ of $\mathbb{R}^{2N-2}$ into $\mathbb{R}^{2N}$ such that $\Gamma$ lies over $\Sigma$, i.e., $\pi\Gamma = \Sigma$ with $\pi : (x,k) \mapsto x$ the canonical projection \cite{Ht}, are of the form $\Gamma = \{(x,k) : x^N=0, k_N=H(\bs{x},\bs{k})\}$ with $H(\bs{x},\bs{k})$ a generic smooth function; correspondingly, the restriction of a symbol is readily defined as $A|_\Gamma (\bs{x},\bs{k})= A\big(\bs{x},0,\bs{k},H(\bs{x},\bs{k})\big)$. Then the boundary value conditions read
\begin{subequations}
\label{6n}
\begin{equation}
\label{6na}
\int\frac{d^Nk}{(2\pi)^N} W(x,k) A(x,k) \Big|_{\Sigma} = \int\frac{d^{N-1}\bs{k}}{(2\pi)^{N-1}}\ W_0 (\bs{x},\bs{k}) A|_\Gamma(\bs{x},\bs{k}),
\end{equation}
which is equivalent to
\begin{equation}
\label{6nb}
W\big|_{\Sigma}(\bs{x},\bs{k},k_N) = 2\pi W_0(\bs{x},\bs{k})\ \delta\big(k_N -H(\bs{x},\bs{k}) \big). 
\end{equation}
\end{subequations}
The function $H$ cannot be arbitrary as \eref{6nb} should satisfy the constraint \eref{1.1a} which reads
\begin{equation}
\label{7n}
D'\big(\bs{x},0,\bs{k}, H(\bs{x},\bs{k})\big)=0,
\end{equation}
hence, the appropriate functions $H$ are obtained on solving the so-called \emph{local dispersion relation} evaluated on $\Sigma$. Since $\Sigma$ has been assumed to be noncharacteristic, i.e., $\partial D'/\partial k_N \not=0$ on $\Gamma$, the function $H$ is well defined and smooth at least locally, in view of the implicit function theorem. On the other hand, it is not unique since \eref{7n} may have multiple solutions, each one corresponding to a specific branch of the dispersion relation. In virtue of the superposition principle for linear wave equations, the total wavefield is a linear superposition of the contributions from each branch of the dispersion relation, the coefficients being determined by the Cauchy boundary values of the normal derivatives $\partial^n \psi/\partial (x^N)^n|_\Sigma$. Therefore, one has a specific Cauchy boundary value problem for the Wigner function of each branch and the sum over all branches yields the total Wigner function. Since one has
\begin{equation*}
|\psi|^2 = \sum_{b,b'} \psi^*_{b'} \psi_b = \sum_b |\psi_b|^2 + \sum_{b\not=b'} \psi^*_{b'}\psi_b,
\end{equation*}
where the indices $b$ and $b'$ run over all branches, and, on noting that the average $\langle \cdot\rangle$ over short scale oscillations cancels out the mixed terms, $\langle\psi_{b'}^*\psi_b\rangle =0$ for $b'\not=b$, whereas $\langle |\psi_b|^2\rangle = |\psi_b|^2$, one gets
\begin{equation}
\label{7nprime}
\langle|\psi(x)|^2\rangle = \sum_b \int \frac{d^Nk}{(2\pi)^N}\ W_b(x,k) = \int \frac{d^Nk}{(2\pi)^N}\ W(x,k),
\end{equation}
that is, the projection of the total Wigner function yields the \emph{averaged wavefield intensity} and, thus, it does not account for, e.g., the formation of short-scale diffraction patterns.

\subsection{The complex geometrical optics method}
Let us now turn to the complex geometrical optics (CGO) method and, in particular, let us discuss its application to pseudodifferential wave equations. This is based on approximating the solution of \eref{1n} by a smooth wave function of the form
\begin{equation}
\label{1.3}
\psi(x) = u(x)\ e^{i\bar{S}(x)} = u(x)\ e^{-\phi(x)} e^{iS(x)}, 
\end{equation}
where, according to the semiclassical limit, $|\partial_xS(x)L| \sim \kappa$ and $\partial_x u(x)| \sim |u(x)|/L$. In addition to the standard oscillating exponential $e^{iS(x)}$, the wave object \eref{1.3} exhibits a novel scale length $w \sim |\partial_x\phi(x)|^{-1}$ which accounts for intermediate-scales variations of the amplitude profile $A(x) = u(x)e^{-\phi(x)}$ with $\kappa \gg |kw| \gg 1$. In general, such an intermediate scalelength $w$ can be determined  by both (strong) absorption \cite{BOG} and diffraction \cite{Mazz}; however, in this paper, it is assumed that the medium is weakly nondissipative [cf. comments after equations \eref{1.1}] so that only diffraction effects are significant. The total short- and intermediate-scale variations of the wavefield are accounted for by the \emph{complex eikonal function} $\bar{S}(x) = S(x) + i\phi(x)$.

The relevant equations for the three unknown functions $u$, $\phi$ and $S$ are determined on substituting the ansatz \eref{1.3} into the wave equation \eref{1n}. For the specific case for which $\hat{D}$ is a differential operator this is straightforward. On the other hand, for the general case, one should deal with the nonlocal response of the operator \cite{MB}. With this aim it is convenient writing \eref{1n} in the configuration space in terms of the Schwartz kernel, namely, 
\begin{equation*}
\int \mathscr{D}(x,x') \psi(x')d^Nx' \sim \int \mathscr{D}^{(s)}\big(\tfrac{x+x'}{2},x-x'\big)\psi(x') d^Nx' = 0,
\end{equation*}
where the exact kernel has been replaced by the $\mathscr{D}^{(s)} = \sigma^{W -1}(D)$ where $D$ is the Weyl symbol. Actually, one could make use of other symbol maps \cite{Mar}, yielding asymptotically equivalent results; here the Weyl-symbol maps has been chosen for direct comparison with the Wigner-Weyl approach. 

Let us further assume that the kernel $\mathscr{D}^{(s)} (x_1,x_2)$ amounts to a distribution smoothly dependent on $x_1$ and with compact support in $x_2$. In virtue of the Paley-Wiener-Schwartz theorem \cite{Ht}, this is equivalent to assume that the corresponding Weyl symbol $D(x,k)$ extends to an \emph{entire} function of the complex-valued dual vector $\bar{k}=k+ik''$, smoothly dependent on $x\in \mathbb{R}^N$. From a physical standpoint, the foregoing assumption implies that nonlocal effects have a finite range: the response $\big(\hat{D}\psi\big)(x)$ of the operator $\hat{D}$ depends only on the value of the wavefield $\psi$ in a compact set. 

Within this condition, one can substitute the complex eikonal ansatz \eref{1.3} into \eref{1n} and expand in Taylor series with respect to $s=x-x'$. As a result one has \cite{MB}
\begin{multline*}
D\big(x,\bar{k}(x)\big) u(x) -i\frac{\partial D\big(x,\bar{k}(x)\big)}{\partial k_{i}}\frac{\partial u(x)}{\partial x^{i}}\\ -\frac{i}{2}\frac{\partial^{2}D\big(x,\bar{k}(x)\big)}{\partial x^{i}\partial k_{i}} u(x) - \frac{i}{2}\frac{\partial^{2} \bar{S}(x)}{\partial x^{i}\partial x^{j}}\frac{\partial^{2} D\big(x,\bar{k}(x)\big)}{\partial k_{i}\partial k_{j}} u(x) = O(\kappa^{-2}),
\end{multline*}
where $\bar{k}(x) = \partial_x S(x) + i\partial_x\phi(x)$ and, for the Paley-Wiener-Schwartz theorem, the estimate $|\partial_x^{\alpha}\partial_k^{\beta} D| = O(\kappa^{m-|\beta|})$ is still valid for complex-extended symbols. To leading orders in $\kappa$, and for a weakly dissipative media, i.e., $D''/D' = O(\kappa^{-1})$, one gets
\begin{subequations}
\label{11n}
\begin{align}
\label{11na}
&D'\big(x, \partial_x\bar{S}(x)\big) = 0,\\
\label{11nb}
&\frac{\partial D'\big(x,\partial_x\bar{S}(x)\big)}{\partial k_i} \frac{\partial u(x)}{\partial x^i}  = \left[D''\big(x,\partial_x\bar{S}(x)\big) - \frac{1}{2}\frac{\partial}{\partial x^i}\Big[\frac{\partial D'\big(x,\partial_x\bar{S}(x)\big) }{\partial k_i}\Big] \right] u(x),
\end{align}   
\end{subequations}
$D'(x, \bar{k})$ and $D''(x,\bar{k})$ being the real and imaginary parts of $D$ extended in the complex $\bar{k}$ space and, in general, are complex valued. It is worth noting that the foregoing equations can be formally obtained from the standard geometrical optics equations, e.g., in the form given by Littlejohn and Flynn \cite{LF}, by replacing $k(x)$ with $\bar{k}(x)$.  

The CGO equations \eref{11n} have been dealt with both by means of the characteristics method in the complex domain \cite{AKF} and on expanding the equations with respect to $\epsilon \equiv |k''(x)|/|k(x)| \sim |k(x)w|^{-1} \ll 1$ \cite{Mazz}. In particular, on referring to the latter approach, in the weak-diffraction regime $\epsilon \sim \kappa^{-1/2}$, terms up to order $\epsilon^{2} \sim \kappa^{-1}$ should be considered in the CGO equation for the complex eikonal $\bar{S}(x)$, which, after separating the real and imaginary parts, amounts to
\begin{subequations}
\label{1.5}
\begin{align}
\label{1.5a}
&D'\big(x,k(x)\big) - \frac{1}{2}\ k''_{i}(x)k''_{j}(x)\ \frac{\partial^{2} D'\big(x,k(x)\big)}{\partial k_{i}\partial k_{j}} = 0,\\
\label{1.5b}
&k''_{i}(x)\ \frac{\partial D'\big(x,k(x)\big)}{\partial k_{i}} = 0.
\end{align}
Equations \eref{1.5a} and \eref{1.5b} constitute a set of coupled first-order partial differential equations for $S(x)$ and $\phi(x)$ with $k(x)=\partial_{x}S(x)$ and $k''(x)=\partial_{x}\phi(x)$.
As for the complex amplitude $u(x)$, only the lowest order approximation with respect to $\epsilon$ is significant, so that the real amplitude $|u(x)|$ is decoupled from the phase $\arg [u(x)]$ (not considered hereafter) and determined by means of the transport equation
\begin{equation}
\label{1.5c}
\frac{\partial}{\partial x^{i}}\left[\frac{\partial D' \big(x,k(x)\big)}{\partial k_{i}}\ |u(x)|^{2}\right] = 2D'' \big(x,k(x)\big)\ |u(x)|^{2}.
\end{equation}
\end{subequations}
This is formally the same equation as the geometrical optics transport equation \cite{LF}, but diffraction effects are accounted for through the wavevector-field $k(x)$ which differs from that obtained in the geometrical optics.  
The approximated form \eref{1.5} of the CGO equations is the one used in physical applications. Moreover, in the zero-diffraction regime ($w \gtrsim L$), one has $\epsilon \sim \kappa^{-1}$, thus terms up to first order only should be considered, with the result that equations \eref{1.5a} and \eref{1.5b} are decoupled and the whole set of CGO equations \eref{1.5} reduces to the standard geometrical optics equations, $\phi$ being effectively zero.

Equations \eref{1.5a} and \eref{1.5b} are usually solved by computing the characteristic curves \cite{T, Ht} of \eref{1.5a} with \eref{1.5b} regarded as a constraint with the result that the characteristics curves thus obtained resemble the geometrical optics rays \cite{KO, LF}. Therefore, the appropriate boundary conditions should be enough to determine the initial values of the complex vector $\bar{k}|_\Sigma(\bs{x})$ evaluated on the boundary surface $\Sigma$. 

Such conditions are obtained from the Cauchy data \eref{1nbound} on writing
\begin{equation}
\label{12n}
\psi_0(\bs{x}) = A_0(\bs{x}) e^{iS_0(\bs{x})} = u_0(\bs{x}) e^{-\phi_0(\bs{x})} e^{iS_0(\bs{x})} + O(\epsilon),
\end{equation}
for some functions $u_0$ and $\phi_0$ such that $|\partial_{\bs{x}}u_0|\sim |u_0|/L$ and $|\partial_{\bs{x}} \phi_0| \sim \phi_0/w$. From \eref{12n} one readily gets the value of the component of the complex vector $\bar{k}= k + ik''$ tangent to $\Sigma$, namely, $\bs{k}(\bs{x}) = \partial_{\bs{x}}S_0(\bs{x})$ and $\bs{k}''(\bs{x}) = \partial_{\bs{x}}\phi_0(\bs{x})$. The remaining normal component is obtained on imposing that the CGO equations \eref{1.5} are satisfied on $\Sigma$; this yields two equation for $k_N$ and $k''_N$, \emph{viz.},
\begin{subequations}
\label{13n}
\begin{align}
\label{13na}
& D'\big(\bs{x},0,\bs{k}(\bs{x}), k_N\big) -\frac{1}{2}\sum_{i,j<N} A_{ij}\big(\bs{x},0,\bs{k}(\bs{x}),k_N\big) k''_i(\bs{x})k''_j(\bs{x}) = 0,\\
\label{13nb}
& k''_N(\bs{x}) = -\sum_{i<N} k''_i(\bs{x}) X_i\big(\bs{x}, 0, \bs{k}(\bs{x}),k_N\big),  
\end{align}
\end{subequations}
where $A_{ij}(x,k) \in S^{m-2}$ and $X_i(x,k) \in S^{0}$ are obtained in terms of the first- and second-order $k$ derivatives of $D'$ and evaluated at $x^N=0$. Equation \eref{13na} is an $O(\epsilon^2)$ perturbation of the local dispersion relation \eref{7n}, hence, it can be solved by
\begin{equation}
\label{14n}
k_N(\bs{x}) = H\big(\bs{x},\bs{k}(\bs{x})\big) + O(\epsilon^2),
\end{equation}
and, in correspondence of \eref{14n}, Eq.\eref{13nb} yields $k''_N(\bs{x})$. As in the Wigner-Weyl formalism, if multiple solutions are found, one should write the wavefield as a sum of contributions from each branch of the local dispersion relation. 

From the foregoing discussion, one should note that the CGO method yields the solution directly in the configuration space, but one should deal with the set of partial differential equations \eref{1.5}, the numerical solution of which can be rather cumbersome. Although the characteristics technique can be used for Eq.\eref{1.5a}, the constraint \eref{1.5b} should be solved in parallel, thus increasing the computational complexity of the problem. As for the global properties of the CGO solution, to our knowledge no general result is still available, though numerical solutions \cite{Mazz} show that the CGO solution is regular even near focal points where the standard geometrical optics solution exhibits a caustic singularity.

In contrast, the Wigner-Weyl formalism appears better suited for numerical solutions. In particular, the wave kinetic equation can be solved along the corresponding Hamiltonian orbits in the phase space so that it is reduced to a set of ordinary differential equations that require limited computational efforts and the solution thus obtained has a global validity in the phase space since the Hamiltonian orbits do not cross each other. In this respect, the constraint \eref{1.1a} does not constitute a limitation as $D'$ is a constant of motion. Moreover, there is no limitation on the nonlocal response of pseudodifferential operators to which the Wigner-Weyl formalism applies. On the other hand, the solution in the phase space should be projected into the configuration space and, thus, an integral with respect to the momentum $k$ should be carried out numerically.

Notwithstanding these differences, the Wigner-Weyl kinetic formalism and the complex geometrical optics method share a number of features, e.g., the solution of the local dispersion relation \eref{14n} relevant to the CGO method is obtained, to the lowest significant order in $\epsilon$, on evaluating the corresponding solution \eref{7n} for $\bs{k} = \bs{k}(\bs{x})$. In the following sections, it will be proved that one can project the wave kinetic equation from the phase space into the configuration space in such a way that the CGO Eq.\eref{1.5} are recovered.

\section{A novel class of solutions to the wave kinetic equation}

As discussed in Sec.2, the solutions of the wave kinetic equation are usually sought in the space $\mathcal{S}'(\mathbb{R}^{2N})$ of tempered distributions \cite{SMM}, or in the space of semiclassical measures \cite{Mar}. The first formulation is the more general, whereas the second follows from the physical requirement that expectation values \eref{star} are well defined. Moreover, we have pointed out that, for general wave propagation problems, physics requires a stronger condition on the Wigner function, namely, the expectation values of physical quantities should be locally defined according to \eref{starstar}. For instance, if one deals with a time-dependent wavefield for which $x^N=ct$ and $k_N = -\omega/c$, the integral
\begin{equation*}
J(\bs{x},t) = \int\frac{d^{N-1}\bs{k}}{(2\pi)^{N-1}}\int\frac{d\omega}{2\pi}\ \frac{\partial D'}{\partial \omega} W, 
\end{equation*}
yields the wave action density $J(\bs{x},t)$ in the space-time \cite{McD}.

In this section, a mathematical characterization of such novel solutions is given and the corresponding differential calculus is put forward.

First, let us note that for any Schwartz function $\varphi (k) \in \mathcal{S}(\mathbb{R}^N)$ and for any tempered distribution $f \in\mathcal{S}'(\mathbb{R}^{2N})$ one can define a tempered distribution $f_{\varphi} \in \mathcal{S}'(\mathbb{R}^N)$ over the configuration space only, given by
\begin{equation}
\label{1mprimo}
\langle f_{\varphi}, \chi\rangle = \langle f, \chi\varphi\rangle =\int f(x,k)\chi(x)\varphi(k) d^Nxd^Nk, \quad \chi \in \mathcal{S}(\mathbb{R}^N),
\end{equation}
where, in general, angle brackets and the integral are alternative ways to denote the action of a distribution on the corresponding test function. The distribution $f$ is smooth with respect to $x\in\mathbb{R}^N$ if and only if $f_\varphi$ amounts to a smooth function $f_\varphi(x)$.
In this case the map $\varphi \mapsto f_\varphi (x)$ for $\varphi \in \mathcal{S}(\mathbb{R}^N)$ defines at every point location $x\in \mathbb{R}^N$ a tempered distribution $f_x \in \mathcal{S}'(\mathbb{R}^{N})$ with
\begin{equation}
\label{1msecondo}
\langle f_x,\varphi\rangle = f_\varphi (x) \qquad \text{for $\varphi \in \mathcal{S}(\mathbb{R}^N)$}.
\end{equation}
This is a consequence of the completeness of $\mathcal{S}'(\mathbb{R}^N)$ along with the identity $f_\varphi ( x) = \lim\limits_{\delta \to 0} \langle f, \chi_\delta (\cdot -x) \varphi\rangle$ where $\chi_\delta$ is a compact-supported function that approximates the Dirac's $\delta$-function for $\delta \to 0$.

Let us now consider a symbol $A\in S^{-\infty} = \cap_m S^m $, that is, $A$ fulfills the symbol estimate \eref{2n} for every order $m$. It follows that $A_x = A(x,\cdot) \in \mathcal{S}(\mathbb{R}^N)$ and for any $f\in\mathcal{S}'(\mathbb{R}^{2N})$ which is smooth with respect to $x$ one can define
\begin{equation}
\label{1m}
\int f(x,k)A(x,k) d^Nk =\langle f_x,A_x\rangle,
\end{equation}
and this is a smooth function on $\mathbb{R}^N$ as required in \eref{starstar}. The definition \eref{1m} should be extended for any symbol $A\in S^m$ with arbitrary order. With this aim, the class of physically admissible solutions [in the sense of \eref{starstar}] is restricted. In particular, it is appropriate considering the tempered distributions $f \in \mathcal{S}'(\mathbb{R}^{2N})$ that satisfy the following conditions:
\begin{enumerate}
\item $f$ is smooth with respect to $x$,
\item the restriction $f_x$ amounts to a distribution with compact support, i.e., $f_x \in \mathcal{E}'(\mathbb{R}^N)$, where $\mathcal{E}'(\mathbb{R}^N)$ is continuously embedded in $\mathcal{S}'(\mathbb{R}^N)$ in the weak topology.
\end{enumerate}
Such a distribution $f$ will be called \emph{momentum distribution} since for every $x$ its restriction $f_x$ represents the distribution of momentum $k$ over $x$. For short let us write $f\in \mathcal{M}$ for the space of momentum distributions.

Within this formulation for every $f\in\mathcal{M}$ and for every $A\in S^m$ Eq.\eref{1m} is well posed and defines a smooth function on $\mathbb{R}^N$. Let us note that the foregoing definition of the space $\mathcal{M}$ is not the optimal one as  functions rapidly decreasing in $k$ are also admissible momentum distributions. However, in the semiclassical limit these functions can be ignored and only compact-supported distributions are significant.

Let us now address the derivatives of a momentum distribution $f\in\mathcal{M}$. First, the derivatives with respect to the momentum $k$ are defined throughout every order. Specifically, since
\begin{equation*}
\langle (\partial^\alpha_k f)_\varphi, \chi\rangle = \langle \partial^\alpha_k f,\chi\varphi\rangle = (-1)^{|\alpha|} \langle f, \chi\partial_k^\alpha \phi\rangle = (-1)^{|\alpha|} \langle f_{\partial_k^\alpha \varphi}, \chi\rangle, 
\end{equation*}
$\partial_k^\alpha f$ is smooth with respect to $x$; moreover, for every $\varphi \in \mathcal{S}(\mathbb{R}^N)$,
\begin{equation*}
\langle (\partial_k^\alpha f)_x, \varphi\rangle = (-1)^{|\alpha|} f_{\partial_k^\alpha \varphi}(x) = (-1)^{|\alpha|} \langle f_x, \partial_k^\alpha \varphi\rangle = \langle \partial_k^\alpha f_x, \varphi\rangle,
\end{equation*}
hence, $(\partial_k^\alpha f)_x$ is compactly supported, and, thus, $\partial_k^\alpha f \in\mathcal{M}$. In terms of the integral notation the latter result evaluated for symbols reads
\begin{equation}
\label{3m}
\int \partial^\alpha_kf(x,k)A(x,k) d^Nk = (-1)^{|\alpha|} \int f(x,k) \partial_k^\alpha A(x,k) d^Nk,
\end{equation}
which is the ``integration-by-parts'' formula. On the other hand, the derivatives with respect to $x$ should be dealt with more carefully. For simplicity, let us consider first-order derivative $\partial f/\partial x^i$ for $f\in \mathcal{M}$. One has that
\begin{equation*}
(\partial f/\partial x^i)_\varphi = \partial f_\varphi / \partial x^i,
\end{equation*}
in virtue of \eref{1mprimo}, so that $\partial f/\partial x^i$ is smooth with respect to $x$. Furthermore,
\begin{equation*}
\langle (\partial f/\partial x^i)_x,\varphi\rangle = \frac{\partial}{\partial x^i} \langle f_x , \varphi\rangle,
\end{equation*}
hence $(\partial f/\partial x^i)_x$ is compactly supported and $\partial f/\partial x^i \in \mathcal{M}$. The explicit formula for the derivative is obtained on noting that for every symbol $A\in S^m$, $Af \in \mathcal{M}$ and
\begin{align*}
\langle (\partial Af /\partial x^i)_x, 1\rangle &= \langle \big(f\partial A/\partial x^i  + A\partial f/\partial x^i \big)_x , 1\rangle \\
&= \langle f_x, (\partial A/\partial x^i)_x\rangle + \langle (\partial f/\partial x^i)_x, A_x\rangle,
\end{align*}    
which in the integral notation takes the form
\begin{equation}
\label{5m}
\int\frac{\partial f(x,k)}{\partial x^i} A(x,k) d^Nk = \frac{\partial }{\partial x^i} \int f(x,k) A(x,k) d^Nk - \int f(x,k) \frac{\partial A(x,k)}{\partial x^i}d^Nk.
\end{equation}
The same result would be obtained from the definition $\lim\limits_{\delta \to 0} \langle \partial f/\partial x^i, \chi_\delta (\cdot - x)A_x \rangle$ through straightforward but longer calculations.
Higher-order derivatives can be defined by recurrence, but they are not explicitly needed in the following.   
   
Searching for solutions of the wave kinetic equations \eref{1.1} in the space of momentum distributions leads to the weak formulation
\begin{subequations}
\label{2.1bar}
\begin{gather}
\label{2.1abar}
\int D'(x,k) W(x,k)\ A(x,k)\ d^{N}k=0,\\
\label{2.1bbar}
\frac{1}{(2\pi)^{N}}\int \Big[\big\{W(x,k),D'(x,k)\big\} - 2 D''(x,k) W(x,k)\Big]A(x,k)\ d^{N}k=0,
\end{gather}
\end{subequations}  
with $W \in \mathcal{M}$. Furthermore, we are interested in semiclassical solutions for which only large-enough momenta are significant. On recalling that $\langle W_x, 1\rangle = (2\pi)^N|\psi(x)|^2 $, we will search for solution of \eref{2.1bar} in the form
\begin{equation}
\label{2.1}
W(x,k) = (2\pi)^{N}\ f\big(x, k-\partial_{x}S(x)\big)\ |\psi(x)|^{2} 
\end{equation}
where $S(x)$ and $|\psi(x)|^2$ are smooth functions to be determined; in particular, $S(x)$ defines a Lagrangian manifold $k=\partial_xS(x)$ in the $x$-$k$ phase space. Moreover, $f \in \mathcal{M}$ is normalized, i.e., $\int fd^Nk = \langle f_x, 1\rangle =1$, and such that
\begin{equation}
\label{2.3}
K_{\alpha}(x) \equiv \int  f(x,\tilde{k}) \tilde{k}^{\alpha} d^{N}\tilde{k}= O(\tilde{w} ^{-|\alpha|}),
\end{equation}
for any multi-index $\alpha$. The integrals in \eref{2.3} are well posed since $\tilde{k}^\alpha = \big(k-\partial_xS(x)\big)^\alpha$ are symbols of order $|\alpha|$, and the corresponding quantities $K_\alpha(x)$ express the statistical moments of the distribution $f$, i.e., they expresses how important are the \emph{deviations} $\tilde{k}$ of the momentum from the Lagrangian manifold $k=\partial_xS(x)$. In particular, $K_0(x)=1$ in view of the normalization condition. In the semiclassical limit it is assumed that the scale length $\tilde{w}$ characterizing the range of the momentum deviations is large enough as compared to $|\partial_xS(x)|^{-1} = |k(x)|^{-1}$, namely, $|k(x)\tilde{w}| \gg 1$.

In Appendix, it is proved that, within the weak formulation \eref{2.1bar}, the momentum distribution $f$, satisfying the foregoing conditions, can be represented by the asymptotic series, cf. Eq.\eref{A.5},
\begin{equation}
\label{2.5}
f\big(x,k-\partial_{x}S(x)\big) \sim \sum_{\beta} \frac{(-1)^{|\beta|}}{\beta!}\ K_{\beta}(x) \partial^{\beta}_{k} \delta\big(k-\partial_{x}S(x)\big),
\end{equation}
controlled by the small parameter $\tilde{\epsilon} \equiv |k(x)\tilde{w}|^{-1} \ll 1$. It is worth noting that $f(x, k-\partial_{x}S(x)\big)$ is thus represented by a distribution which is point supported on the Lagrangian manifold $k=\partial_{x}S(x)$ and completely determined by its statistical moments $K_{\beta}(x)$.

In correspondence of the asymptotic expansion \eref{2.5}, Eq.\eref{2.1abar} reduces to
\begin{equation}
\label{2.6}
\sum_{\alpha}\ \frac{1}{\alpha!}\ \partial_{k}^{\alpha} D' \big(x,\partial_{x}S(x)\big)\ K_{\alpha + \beta} (x) = 0,
\end{equation} 
as shown in details in Appendix, cf., in particular, Eq.\eref{A.6}. Formally, Eq.\eref{2.6} constitutes an infinite set of algebraic equations for the statistical moments $K_{\alpha}(x)$ characterizing the momentum distribution, where each equation, labelled by $\beta$, is expressed as an asymptotic series in $\tilde{\epsilon}$; the function $S(x)$ is determined by imposing that the system \eref{2.6} admits nontrivial solutions. Equations \eref{2.6} are valid for a general momentum distribution $f$ which satisfies \eref{2.1abar}. In particular, on setting $K_{\alpha}(x)=0$ for $\alpha \not =0$, Eq.\eref{2.5} reduces to
\begin{equation*}
f\big(x,k-\partial_xS(x)\big) = (2\pi)^N \delta\big(k-\partial_xS(x)\big), 
\end{equation*}
which is the geometrical-optics-like solution obtained by Bornatici and Kravtsov \cite{BK} and by Sparber, Markowich and Mauser \cite{SMM}, whereas in \eref{2.6} the only nontrivial equation reduces to the geometrical optics eikonal equation \cite{LF} for $S(x)$, namely, $D'\big(x,\partial_xS(x)\big)=0$.

\section{The CGO-like solution of the wave kinetic equation}

On the basis of the asymptotic expansion \eref{2.5} for a momentum distribution, one can prove the main result of this paper, that is, relating the wave kinetic equation to the CGO equations for suitable boundary conditions.

First let us consider the specific momentum distribution for which
\begin{equation}
\label{2.7}
K_{\alpha}(x) = \begin{cases}
0,\qquad\qquad\qquad\;\;\; \text{for $|\alpha| = 2n + 1$ (odd),}\\
(-1)^{n} \big(k''(x)\big)^{\alpha}, \quad \text{for $|\alpha| = 2n$ (even),}
\end{cases}
\end{equation}
that is, odd-order moments have been set to zero, whereas even-order moments have been related to a single vector field $k''(x)=\partial_{x}\phi(x)$, with $\phi$ an unknown smooth function. Correspondingly, the momentum distribution \eref{2.5} takes the form 
\begin{align}
\nonumber
f\big(x, k-\partial_{x}S(x)\big) &\sim \sum_{|\beta|=\text{even}} \frac{(-1)^{\frac{|\beta|}{2}}}{\beta!}\ \big(k''(x)\big)^{\beta} \partial_{k}^{\beta}\delta\big(k-\partial_{x}S(x)\big) \\
\label{2.8bar}
&= \sum_{n=0}^{+\infty} \frac{(-1)^{n}}{(2n)!} \left[k''_{i}(x) \frac{\partial}{\partial k_{i}} \right]^{2n} \delta\big(k-\partial_{x}S(x)\big),
\end{align}
the second identity being obtained by means of the multinomial formula
\begin{equation}
\label{2.9bar}
\sum_{|\beta|=n} \frac{1}{\beta!}\ a_{1}^{\beta_{1}}\cdots a_{N}^{\beta_{N}} = \frac{1}{n!} \big(a_{1} + \cdots + a_{N}\big)^{n},
\end{equation}
with $a_{i}=k''_{i} \partial /\partial k_{i}$ (no sum over $i$). Let us note that the momentum distribution \eref{2.8bar} is symmetric with respect to the Lagrangian manifold $k=\partial_{x}S(x)$, as even-order moments only appear; in particular, the second order moment $-k''_{i}(x)k''_{j}(x)$ for $i=j$ is negative, so that such a distribution cannot be interpreted as a probability measure.

The momentum distribution \eref{2.8bar} should be multiplied by $|\psi(x)|^2$ to get the whole Wigner function \eref{2.1}. Let us consider the case for which
\begin{equation}
\label{2.10bar}
|\psi(x)|^{2}= |u(x)|^{2}\ e^{-2\phi(x)}, 
\end{equation}
$\phi(x)$ being defined in \eref{2.7} and $|u(x)|^2$ is ordered according to $|\partial_x u|\sim |u|/L$ with $L\gg \tilde{w} \sim |\partial_x\phi|^{-1}$. As a consequence, $\tilde{w}$ is the shortest scale length characterizing the wavefield intensity $|\psi(x)|^2$, hence it can be identified with the scale length $w$ defined after Eq.\eref{1.3}, namely, $\tilde{w}\sim w$ and $\tilde{\epsilon} \sim \epsilon$.

Then one has the following:

\vspace{5mm}
\noindent\emph{
The Wigner function $W(x,k) = (2\pi)^N f\big(x, k-\partial_xS(x)\big) |\psi(x)|^2$, with $f$ given by \eref{2.8bar} and $|\psi(x)|^2$ expressed in the form \eref{2.10bar}, satisfies asymptotically the wave kinetic equation in the weak formulation \eref{2.1bar} with an $O(\epsilon)$ remainder if and only if 
\begin{itemize}
\item[(i)] the smooth functions $S(x)$ and $\phi(x)$ satisfy the complex geometrical optics equations \eref{1.5a} and \eref{1.5b},
\end{itemize}
and
\begin{itemize}
\item[(ii)] the smooth slowly varying function $|u(x)|^2$ satisfies the transport equation \eref{1.5c}.
\end{itemize} 
}\vspace{5mm}

First, let us prove the statement $(i)$. In view of the ansatz \eref{2.7}, all the equations obtained from \eref{2.6} with $\beta$ such that $|\beta|$ is an even integer reduce to the same equation which reads
\begin{subequations}
\label{2.10}
\begin{equation}
\label{2.10a}
\sum_{|\alpha|=\text{even}} \frac{(-1)^{\frac{|\alpha|}{2}}}{\alpha !}\ \partial^{\alpha}_{k}D'\big(x,k(x)\big) \big(k''(x)\big)^{\alpha} = \sum_{n=0}^{+\infty} \frac{(-1)^{n}}{(2n)!}\ \left[k''_{i}(x)\frac{\partial}{\partial k_{i}}\right]^{2n} D'\big(x,k(x)\big) =0,
\end{equation}
and, analogously, all the equations obtained from \eref{2.6} with $\beta$ such that $|\beta|$ is an odd integer reduce to
\begin{equation}
\label{2.10b}
\sum_{|\alpha|=\text{odd}} \frac{(-1)^{\frac{|\alpha|+1}{2}}}{\alpha !}\ \partial^{\alpha}_{k}D' \big(x,k(x)\big) \big(k''(x)\big)^{\alpha} =\sum_{n=0}^{+\infty} \frac{(-1)^{n+1}}{(2n+1)!}\ \left[k''_{i}(x)\frac{\partial}{\partial k_{i}}\right]^{2n+1} D'\big(x,k(x)\big)=0,
\end{equation}
\end{subequations}
where the second identity in both equations \eref{2.10} follows on using the multinomial formula \eref{2.9bar}. Equations \eref{2.10} constitute a set of two coupled equations for the real functions $S(x)$ and $\phi(x)$, and, to lowest significant orders in $\epsilon$, they are the same as the CGO equations \eref{1.5a} and \eref{1.5b}; this completes the proof of $(i)$. 

As for $(ii)$, on account of the differential calculus for momentum distributions put forward in Sec.3, the term connected with the Poisson brackets in Eq.\eref{2.1bbar} should be written in the form, cf. Eq.\eref{5m}, 
\begin{multline}
\label{2.12}
\int\frac{d^Nk}{(2\pi)^N}\ \big\{W,D'\big\} A = \frac{\partial}{\partial x^{i}} \int\frac{d^Nk}{(2\pi)^N}\ W\frac{\partial D'}{\partial k_{i}} A \\
- \int\frac{d^Nk}{(2\pi)^N}\ W \frac{\partial}{\partial x^{i}}\left(\frac{\partial D'}{\partial k_{i}} A\right) -\int\frac{d^Nk}{(2\pi)^N}\ \frac{\partial W}{\partial k_{i}}\frac{\partial D'}{\partial x^{i}} A.
\end{multline}
Using the specific momentum distribution \eref{2.8bar} for which
\begin{equation}
\label{16bar}
W(x,k) = (2\pi)^N \delta\big(k-\partial_xS(x) \big) |\psi(x)|^2 + O(\epsilon^2)
\end{equation}
yields
\begin{align}
\nonumber
\int\frac{d^Nk}{(2\pi)^N}\ \big\{W,D'\big\} A &= \frac{\partial}{\partial x^i} \left[\frac{\partial D'\big(x,k(x)\big)}{\partial k_i} |\psi(x)|^2 A\big(x, k(x)\big) \right] - \left[\frac{\partial D'\big(x,k(x)\big)}{\partial k_i}\frac{\partial A\big(x,k(x)\big)}{\partial x^i}\right. \\
\nonumber
& \qquad\qquad \left. -\frac{\partial D'\big(x,k(x)\big)}{\partial x^i} \frac{\partial A\big(x,k(x)\big)}{\partial k_i}\right] |\psi(x)|^2 + O(\epsilon^2)\\
\label{17n}
&= \frac{\partial}{\partial x^i} \left[\frac{\partial D'\big(x,k(x)\big)}{\partial k_i} |\psi(x)|^2  \right] A\big(x, k(x)\big) + O(\epsilon).
\end{align}
The last identity follows on noting that taking the derivative of \eref{2.10a} with respect to $x^{j}$ yields
\begin{equation*}
\frac{\partial D' \big(x,k(x)\big)}{\partial k_{i}}\frac{\partial k_{j}(x)}{\partial x^{i}} = \frac{\partial D' \big(x,k(x)\big)}{\partial k_{i}}\frac{\partial k_{i}(x)}{\partial x^{j}} = - \frac{\partial D' \big(x,k(x)\big)}{\partial x^{j}} + O(\epsilon)
\end{equation*}
with $k_{j}(x)=\partial S(x) / \partial x^{j}$. Equation \eref{17n} implies that
\begin{equation*}
\big\{W,D' \big\}= (2\pi)^{N}\frac{\partial}{\partial x^i} \left[\frac{\partial D'\big(x,k(x)\big)}{\partial k_i} |\psi(x)|^2  \right] \delta\big(k-\partial_{x}S(x)\big)  + O(\epsilon)
\end{equation*}
in the weak sense. Hence, from the wave kinetic equation, to lowest order in $\epsilon$, one gets the transport equation
\begin{equation*}
\frac{\partial}{\partial x^{i}}\left[\frac{\partial D'\big(x,k(x)\big)}{\partial k_{i}}\ |\psi(x)|^{2}\right] =  2 D''\big(x, k(x)\big)\ |\psi(x)|^{2},
\end{equation*}
which reduces to the CGO transport Eq.\eref{1.5c} for $|u(x)|^{2}$, cf. equation \eref{2.10bar}, on noting that, to lowest significant order,
\begin{equation*}
\frac{\partial}{\partial x^{i}}\left[\frac{\partial D'\big(x, k(x)\big)}{\partial k_{i}}\ |u(x)|^{2} e^{-2\phi(x)}\right] =  e^{-2\phi(x)}\ \frac{\partial}{\partial x^{i}}\left[\frac{\partial D'\big(x,k(x)\big)}{\partial k_{i}}\ |u(x)|^{2}\right],
\end{equation*}
in view of Eq.\eref{2.10b}. This concludes the proof. 

The foregoing result shows that there exists a specific form of the Wigner function for which the wave kinetic equation is reduced to the CGO equations.

In order to compare the wavefield intensities predicted by the wave kinetic description with that obtained on solving the CGO equations, one should complete the foregoing argument by discussing Cauchy boundary conditions. With reference to \eref{6n} and \eref{7n} one has the following:

\vspace{5mm}
\noindent\emph{%
Let $W(x,k)$ be the weak solution of the wave kinetic equation \eref{1.1} corresponding to the Cauchy boundary conditions 
\begin{equation}
\label{20n}
W\big|_{\Sigma}(\bs{x},\bs{k},k_N) = (2\pi)^N \delta\big(\bs{k}-\bs{k}(\bs{x})\big) \delta\big(k_N - H(\bs{x},\bs{k}(\bs{x}))\big) |u_0(\bs{x})|^2 e^{-2\phi_0(\bs{x})} + O(\epsilon),
\end{equation}
for some smooth functions $S_0(\bs{x})$, $\phi_0(\bs{x}) \geq 0$ and $|u_0(\bs{x})|^2$ with $\bs{k}(\bs{x})=\partial_{\bs{x}}S_0(\bs{x})$ satisfying the CGO ordering defined after \eref{12n}, and let $S(x)$, $\phi(x)$ and $|u(x)|^2$ be solution of the CGO equations \eref{1.5} with Cauchy boundary conditions given by the same function $S_0$, $\phi_0$ and $|u_0|^2$. Then $W(x,k)$ can be approximated according to
\begin{equation}
\label{21n}  
W(x,k) = (2\pi)^N \delta\big(k-\partial_xS(x)\big) |u(x)|^2 e^{-2\phi(x)} + O(\epsilon)
\end{equation}
in the weak sense of Sec.3.
}\vspace{5mm}

First, let us note that the Cauchy data \eref{20n} is a particular case of \eref{6nb} which corresponds to
\begin{equation}
\label{22n}
W_0(\bs{x},\bs{k}) = (2\pi)^{N-1} \delta\big(\bs{k}-\partial_{\bs{x}}S_0(\bs{x})\big) |u_0(\bs{x})|^2 e^{-2\phi_0(\bs{x})} + O(\epsilon);
\end{equation}
in particular, the Wigner function corresponding to the complex-eikonal wave object \eref{12n} can be written in the form \eref{22n}.

In order to prove the foregoing statement, we will make use of the previous result of this section. Specifically, we have proved that the Wigner function 
\begin{equation*}
\tilde{W}(x,k) = (2\pi)^N f\big(x,k-\partial_xS(x)\big) |u(x)|^2 e^{-2\phi(x)},
\end{equation*}
$f$ being the momentum distribution \eref{2.8bar}, solves asymptotically the wave kinetic equation in the weak sense within an $O(\epsilon)$ accuracy. Moreover,
\begin{equation*}
\tilde{W}(x,k) =  (2\pi)^N \delta\big(k-\partial_xS(x)\big) |u(x)|^2 e^{-2\phi(x)} + O(\epsilon^2),
\end{equation*}
in view of \eref{16bar}. As for the boundary condition \eref{20n} one gets
\begin{align*}
\tilde{W}\big|_{\Sigma} (\bs{x},\bs{k},k_N) &= (2\pi)^N \delta\big(\bs{k}-\partial_{\bs{x}}S(\bs{x},0)\big) \delta\big(k_N -\partial_{x^N} S(\bs{x},0)\big) |u(\bs{x},0)|^2 e^{-2\phi(\bs{x},0)}+ O(\epsilon^2)\\
&= (2\pi)^N \delta(\bs{k}-\partial_{\bs{x}}S_0(\bs{x})\big) \delta\big(k_N -k_N(\bs{x})\big) |u_0(\bs{x})|^2 e^{-2\phi_0(\bs{x})} + O(\epsilon^2),
\end{align*}
where $k_N(\bs{x}) = \partial_{\bs{x}}S(\bs{x},0)$. According to \eref{14n}, $k_N(\bs{x}) = H\big(\bs{x},\bs{k}(\bs{x})\big) + O(\epsilon^2)$, and, thus, $\delta\big(k_N-k_N(\bs{x})\big) = \delta\big(k_N - H(\bs{x},\bs{k}(\bs{x}))\big) + O(\epsilon^2)$ in the weak sense, so that $\tilde{W}(x,k)$ matches the boundary conditions \eref{20n}. Since the solution of the wave kinetic equation along with the Cauchy boundary condition \eref{20n} is unique and since $\tilde{W}$ is an $O(\epsilon)$ solution, it follows that
\begin{equation*}
W(x,k) = \tilde{W}(x,k) + O(\epsilon),
\end{equation*}
which concludes the proof of \eref{21n}.

This implies that, whenever the solutions of both the wave kinetic equation and the CGO equations exist, thus, in particular, the Cauchy boundary conditions are of the form \eref{22n}, the Wigner-Weyl formalism and the complex geometrical optics method are equivalent within an $O(\epsilon)$ accuracy. In particular, the wavefield intensity predicted by the Wigner-Weyl kinetic formalism is the same as that predicted by the CGO method, namely,
\begin{equation}
\label{intensity}
|\psi(x)|^2 = \int\frac{d^Nk}{(2\pi)^N} W(x,k) = |u(x)|^2 e^{-2\phi(x)} + O(\epsilon),
\end{equation}
the second identity following from \eref{21n}. In the next section, an analytically tractable case is considered as an example. Specifically, the solution of the wave kinetic equation relevant to the paraxial propagation of a Gaussian wave beam in a ``lens-like'' medium is obtained and shown to be the same as the corresponding CGO solution.

\section{The kinetic description of diffraction effects for a ``lens-like'' medium and its analogy with the quantum harmonic oscillator}

Let us address the case of a monochromatic $(e^{-i\omega t})$ beam of electromagnetic waves propagating in a \emph{loss less} ``lens-like'' medium \cite{BM} with \emph{real} refractive index $n({\bf r},\omega)\equiv n(x)=n_{0}\big[1-(x/L)\big]^{\frac{1}{2}}$. It is assumed that the wavefield is localized near the axis $x=0$ of the medium, that is, $(x/L) \ll 1$; moreover, the wave electric field is written in the form ${\bf E}({\bf r},\omega)={\bf \hat{y}}\ E(x,z;\omega)$, i.e., it is polarized along the $y$ axis and propagates in the $x$-$z$ plane. The relevant wave equation for the wavefield real amplitude $E(x,z;\omega)$ is thus the Helmholtz equation. The corresponding Weyl symbol is real valued and given by
\begin{equation}
D(x, k_{x},k_{z})=-\big(k_{x}^{2} +k_{z}^{2}\big) +\frac{\omega^{2}}{c^{2}} n^{2}(x)=-\big(k_{x}^{2} +k_{z}^{2}\big) +\frac{\omega^{2}}{c^{2}} n^{2}_{0}\Big[1-(x/L)^{2}\Big], 
\end{equation}
thus, the dispersion relation $D=0$ yields two branches, to be referred to as the progressive and the regressive waves. As for the Cauchy boundary conditions, let us assume that the wavefield is purely Gaussian at $z=0$, i.e., $E(x,0;\omega)=u_{0}\exp\big[-(x-x_{0})^{2}/w_{0}^{2}\big]$, $w_{0}$ being the initial width, and the propagation occurs along the $z$ axis, so that one should solve the dispersion relation $D=0$ for $k_{z}$. On assuming that each branch of the dispersion relation carries half of the wavefield intensity, one can consider the progressive wave only which reads
\begin{equation}
\label{3.1}
\frac{k_{z}}{k_{0}} = \sqrt{1- \Big(\frac{x}{L}\Big)^{2} - \Big(\frac{k_{x}}{k_{0}}\Big)^{2}} \simeq 1-\frac{1}{2}\Big(\frac{x}{L}\Big)^{2} - \frac{1}{2}\Big(\frac{k_{x}}{k_{0}}\Big)^{2}
\end{equation}
where $k_{0}=\omega n_{0}/c$ is the wavevector at $x=0$ and the paraxial approximation $(k_{x}/k_{0})^{2} \sim (x/L)^{2} \sim (w/L)^{2} \sim \lambda/L \ll 1$ has been exploited as relevant to the weak-diffraction regime \cite{BM}. It is convenient noting that the dispersion relation corresponding to the second form of \eref{3.1} can be written as
\begin{equation}
 \frac{1}{k_{0}}\big(k_{0}-k_{z}) - \frac{1}{2}\Big(\frac{k_{x}}{k_{0}}\Big)^{2} - \frac{1}{2}\Big(\frac{x}{L}\Big)^{2}  = 0,
\end{equation}
which is formally analogous to the dispersion relation relevant to a quantum harmonic oscillator \cite{McD} with unit mass and $1/k_{0}\rightarrow \hbar$, $1/L \rightarrow \omega_{0}$, $\omega_{0}$ being the characteristic frequency of the oscillator, $z\rightarrow t$ and $k_{0}-k_{z}(>0) \rightarrow \omega$. In particular, the frequency $\omega$ corresponds to the shifted wavevector $k_{0}-k_{z}$ along the propagation direction $z$. The shift occurs because of the oscillations of the wavefield along the propagation direction $z$. 

This analogy allows to make use of the well-known solution of the wave kinetic equation for the quantum harmonic oscillator \cite{McD} to describe the paraxial propagation of a Gaussian beam in the ``lens-like" medium. More specifically, the solution of the wave kinetic equation for the harmonic oscillator corresponding to an initially Gaussian wave packet $\psi(x,0)=(w_{0}\sqrt{\pi/2})^{-1/2}\ \exp\big[-(x-x_{0})^{2}/w_{0}^{2}\big]$ is \cite{McD} 
\begin{subequations}
\label{3.2pre}
\begin{gather}
\label{3.2prea}
|\psi(x,t)|^{2} = \sqrt{\frac{2}{\pi w(t)^{2}}}\ \exp\left(-2 \frac{\big(x-x_{0}\cos(\omega_{0} t)\big)^{2}}{w(t)^{2}}\right),\\
w^{2}(t)=\big[\cos^{2}(\omega_{0} t) +\varepsilon^{2}\sin^{2}(\omega_{0} t)\big]\ w_{0}^{2}, 
\end{gather}
\end{subequations}
where $w(t)$ is the width of the wave packet as a function of time and $\varepsilon=2\hbar/m\omega_{0} w_{0}^{2}$, $m$ being the mass of the oscillator and $x_{0}$ the initial displacement of the Gaussian from the centre of the elastic force acting on the oscillator. 

Correspondingly, the solution for the wave electric field intensity in the ``lens-like'' medium, with the considered launching conditions, is
\begin{subequations}
\label{3.2} 
\begin{gather}
\label{3.2a}
\langle |{\bf E}(x,z; \omega)|^{2}\rangle = u_{0}^{2}\ \frac{w_{0}}{w(z)}\ \exp\left(-2 \frac{\big(x-x_{0}\cos(z/L)\big)^{2}}{w(z)^{2}}\right),\\
\label{3.2b}
w^{2}(z)=\left[\cos^{2}(z/L) +\Big(\frac{2L}{k_{0}w_{0}^{2}}\Big)^{2}\sin^{2}(z/L)\right]w_{0}^{2} =  \left[1+\Big(\Big(\frac{L}{z_{R}}\Big)^{2}-1\Big)\sin^{2}(z/L)\right] w_{0}^{2},
\end{gather}
\end{subequations}
with $z_{R}= k_{0}w_{0}^{2}/2$ the Rayleigh range in the medium. In Eq.\eref{3.2b}, it has been explicitly indicated that the solution obtained from the wave kinetic equation amounts to the averaged intensity $\langle |{\bf E}(x,z;\omega)|^{2}\rangle$, rather than to the exact value  $|{\bf E}(x,z;\omega)|^{2}$ since two branches of the dispersion relation exist each one carrying half of the wavefield intensity, cf. comments after Eqs.\eref{7nprime}. The intensity \eref{3.2a} and the beam width \eref{3.2b} are the same as the corresponding quantities obtained from the CGO solution \cite{BM}. As a consequence the intensity profile \eref{3.2} accounts for diffraction effects as shown in Fig.1. 
\begin{figure}
\includegraphics[bb =54 540 539 705, scale = .8]{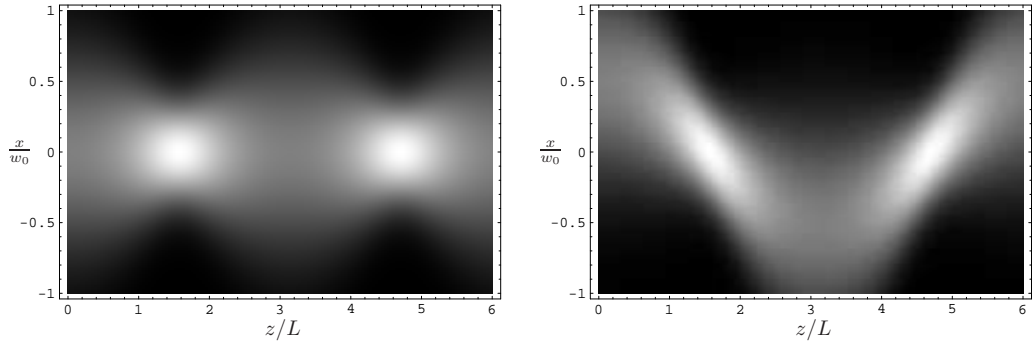}
\caption{The wavefield intensity \eref{3.2} in the $(x,z)$-plane as obtained from the wave kinetic equations \eref{1.1} (bright regions correspond to high intensity), for the case $L/z_{R}=0.5$ [cf. equation \eref{3.2b}] with $x_{0}=0$ and $x_{0}=\frac{1}{2}w_{0}$, respectively. One should note that the wave beam exhibits a finite width even near focal points (characterized by bright spots) where the geometrical optics solution exhibits caustic singularities.}
\end{figure}

One can conclude that, according to results of Sec.4, the \emph{kinetic formalism can be used to describe the effects of diffraction on the propagation of wave beams}, and, for the case under consideration, it yields the same result as the CGO method. Nevertheless, the detailed structure of the wavefield, i.e., the oscillations along the propagation direction $z$, the effects of the curvature of phase fronts and the Gouy shift, which are available from the CGO solution \cite{BM}, \emph{cannot be resolved by means of the wave kinetic equation}, which instead gives the averaged intensity distribution.   
  
\section{Conclusions}

Within the framework of semiclassical wave propagation, two specific asymptotic techniques have been considered, namely, the Wigner-Weyl kinetic formalism and the complex geometrical optics (CGO) method. A detailed comparative analysis of these techniques has been given in Sec.2, for the case of scalar pseudodifferential wave equations, with Cauchy boundary conditions. 

In particular, in the Wigner-Weyl formalism, the wavefield is represented in the phase space by the Wigner function which is a solution to the wave kinetic equation. In the most general case the Wigner function amounts to a tempered distribution. However, physical considerations lead to the definition of a novel class of weak solutions which have been characterized in Sec.3. Such specific weak solutions are referred to as \emph{momentum distributions} since, for every point location $x$ in the configuration space, they give the distribution of momentum $k$ over $x$, in the $x$-$k$ phase space.

On the other hand, the CGO method yields an asymptotics solution of a pseudodifferential wave equation directly in the configuration space, in terms of three smoothly varying functions representing, the phase, the wavefield envelope and the amplitude, respectively.

In Sec.4, on the basis of the mathematical framework developed in Sec.3, we have proved that, whenever both the solutions of the wave kinetic equation and of the CGO equations exist, thus, in particular, the Cauchy data are of the form \eref{22n}, the former can be approximated by a momentum distribution, cf., equation \eref{21n}, written in terms of the three smoothly varying functions that solve the CGO equations \eref{1.5}. As a consequence, the two considered techniques are asymptotically equivalent and, in particular, to lowest significant order, the wavefield intensity predicted by the Wigner-Weyl formalism is the same as that predicted by the CGO method, cf. Eq.\eref{intensity}. 

In addition, one can conclude that the Wigner-Weyl kinetic formalism properly describes the wavefield near focal points. This is also shown by comparing the solution of the kinetic equation to that of the CGO equations for a specific case, namely, the propagation of electromagnetic Gaussian wave beams in an isotropic ``lens-like'' medium, cf. Sec.5. In particular, the relevant solution of the wave kinetic equation has been obtained on the basis of the analogy between the ``lens-like'' medium and the quantum harmonic oscillator.

\section*{acknowledgments}
The author is grateful to M.~Bornatici for many useful discussions, suggestions and for carefully reading the manuscript. 
For the revised version of this paper, valuable discussions with C.~Dappiaggi and A.~Orlandi are gratefully acknowledged. 
This work was supported by the Italian Ministry of University Scientific Research and Technology (MURST) and the Italian Institute for the Physics of Matter (INFM).

\appendix
\section{The asymptotic series expansion of the momentum distribution}

In this appendix the asymptotic series expansion \eref{2.5} of the momentum distribution is proved and the corresponding equations \eref{2.6} are derived from the weak form \eref{2.1abar} of the dispersion relationship \eref{1.1a}.

Since both $D'(x,k)$ and $A(x,k)$ are symbols, they are, in particular, smooth functions, and one can apply the Taylor's formula 
\begin{equation}
\label{A.2}
\begin{bmatrix}
D'\\ A \end{bmatrix}
\big(x,k(x)+\tilde{k}\big) = \sum_{|\alpha|\leq n-1} \frac{1}{\alpha!}\ 
\begin{bmatrix}
\partial_{k}^{\alpha} D' \\ \partial_{k}^{\alpha}A\end{bmatrix} \big(x,k(x)\big) \tilde{k}^{\alpha} + \sum_{|\alpha|=n} 
\begin{bmatrix}
d_{\alpha}(x,\tilde{k})\\ a_{\alpha}(x,\tilde{k}) \end{bmatrix} \tilde{k}^{\alpha},
\end{equation}
$\alpha=(\alpha_{1},\ldots,\alpha_{N})$ being an $N$-dimensional multi-index and
\begin{equation*}
\begin{bmatrix}
d_{\alpha}(x,\tilde{k})\\ a_{\alpha}(x,\tilde{k}) \end{bmatrix}= \frac{|\alpha|}{\alpha!}\ \int_{0}^{1}dt\ (1-t)^{n-1}\ \begin{bmatrix}
\partial_{k}^{\alpha} D' \\ \partial_{k}^{\alpha}A\end{bmatrix} 
\big(x,(1-t)k(x) + t\tilde{k}\big)
\end{equation*}
the remainder of order $n$ relevant to the expansions of $D'$ and $A$, respectively. More specifically, on making use of \eref{A.2} to evaluate the left-hand side of \eref{2.1abar}, one gets 
\begin{multline}
\label{A.3}
\int  f(x,\tilde{k}) D'\big(x,k(x)+\tilde{k}\big) A\big(x,k(x)+\tilde{k}\big)\ d^{N}\tilde{k}\\
\sim \sum_{\alpha,\beta} \frac{1}{\alpha!\beta!}\ \partial _{k}^{\alpha}D' \big(x,k(x)\big)\partial_{k}^{\beta}A\big(x, k(x))\ K_{\alpha +\beta}(x)
\end{multline}
where $K_{\alpha}(x)=O(\tilde{w}^{-|\alpha|})$ are the statistical moments of the momentum distribution $f(x,\tilde{k})$, c.f. Eq.\eref{2.3}. In virtue of the symbol estimate \eref{2n}, symbols are such that, e.g., $|\partial_{k}^{\alpha} A(x,k)| = O( |k|^{m-|\alpha|})$ in the semiclassical limit $|k| \rightarrow +\infty$ uniformly in $x$, hence, the asymptotic series expansion  \eref{A.3} is controlled by the (small) parameter $\tilde{\epsilon} \equiv |k(x)\tilde{w}|^{-1}$. Moreover, on noting that
\begin{equation*}
\partial^{\beta}_{k}A\big(x,k(x)\big) = (-1)^{|\beta|} \int \partial^{\beta}_{k}\delta\big(k-k(x)\big)\ A(x,k)\ d^{N}k,
\end{equation*}
Eq.\eref{A.3} takes the form
\begin{multline*}
\int D'(x,k) f\big(x,k-k(x)\big) A(x,k)\ d^{N}k  \\
\sim \int \left[\sum_{\alpha, \beta} \frac{(-1)^{|\beta|}}{\alpha!\beta!}\ \partial_{k}^{\alpha}D'\big(x,k(x)\big) K_{\alpha + \beta}(x) \partial_{k}^{\beta}\delta\big(k-k(x)\big)\right] A(x,k)\ d^{N}k
\end{multline*}
and, in view of the arbitrariness of $A(x,k)$, one gets
\begin{equation}
\label{A.4}
D'(x,k) f\big(x,k-k(x)\big) \sim \sum_{\alpha, \beta} \frac{(-1)^{|\beta|}}{\alpha!\beta!}\ \partial_{k}^{\alpha}D'\big(x,k(x)\big) K_{\alpha + \beta}(x) \partial_{k}^{\beta}\delta\big(k-k(x)\big),
\end{equation}
in the weak sense. It is worth noting that the derivation of \eref{A.4} does not depend on the explicit form of the symbol $D'(x,k)$, thus, on setting $D'(x,k)=1$, Eq.\eref{A.4} reduces to
\begin{equation}
\label{A.5} 
f\big(x,k-k(x)\big) \sim \sum_{\beta}\frac{(-1)^{|\beta|}}{\beta!}\ K_{\beta}(x)\partial^{\beta}_{k}\delta\big(k-k(x)\big)
\end{equation}
which is just the general asymptotic expansion \eref{2.5} of the momentum distribution. 

Going back to Eq.\eref{2.1abar}, its solution is obtained on setting the expansion \eref{A.4} to zero and exploiting the linear independence of the derivatives of the Dirac's $\delta$-function, thus yielding a set of equations for the statistical moments, namely,
\begin{equation}
\label{A.6}
\sum_{\alpha}\frac{1}{\alpha!}\ \partial^{\alpha}_{k}D'\big(x,k(x)\big) K_{\alpha + \beta}(x)=0,
\end{equation}
which is just Eq.\eref{2.6}. It is worth noting that Eq.\eref{A.6} can be also obtained on substituting \eref{A.5} into \eref{2.1abar} and exploiting the Leibniz's formula
\begin{equation*}
\partial_{k}^{\beta}\big(D' A\big) = \sum_{\alpha + \gamma =\beta} \frac{\beta!}{\alpha!\gamma!}\ \partial_{k}^{\alpha}D'\partial_{k}^{\gamma}A
\end{equation*}
which expresses the derivative of a product to any orders.

\end{document}